\documentclass[aps,twocolumn,superscriptaddress]{revtex4}
\usepackage{amsmath,amssymb}
\usepackage{graphics,graphicx}
\usepackage{dcolumn,bm}
\usepackage{psfrag}
\usepackage{color}
\newcommand{\cu}
{\affiliation{Department of Physics, University of Calcutta,
92 Acharya Prafulla Chandra Road, Kolkata 700009, India.}}

\begin{document}

\title{Effect of bias in a reaction diffusion system in two dimensions}

\author{Pratik Mullick}%
\cu
\author{Parongama Sen}%
\cu

\begin{abstract}
We consider a single species reaction diffusion system on a two dimensional lattice  where 
the particles $A$   are biased to move towards their nearest neighbours 
and annihilate 
as they meet. Allowing the bias to take both negative and positive values parametrically, any nonzero bias is seen to drastically affect the behaviour of the system compared to the unbiased (simple diffusive) case. 
For positive bias, a finite number of dimers,
which are isolated pairs of 
particles occurring as nearest neighbours, 
exist  while for negative bias, a finite density 
of particles survives.  
Both the quantities vanish in a power law manner close to the diffusive limit 
with different exponents.
The appearance of dimers is exclusively due to the parallel updating scheme used in the simulation.
The results indicate the presence of a continuous phase transition at the diffusive point.
In addition, a discontinuity is observed at the fully positive bias limit.  
The persistence behaviour is also analysed for the system. 
\end{abstract}

\maketitle

\section{Introduction}

Non-equilibrium systems of diffusing particles, which undergo reactions such as 
pairwise annihilation, have received a lot of attention in recent times
\cite{books}.
 The particles 
may represent molecules, 
biological entities, or even abstract quantities like opinions in societies or market commodities.  Such systems are also widely 
used to describe the pattern-formation phenomena in a  variety of biological, chemical 
and physical systems. In a  lattice representation  of the simple single species problem, the  lattice 
sites are  occupied by  a particle  at time $t=0$ with a certain probability. At each time step, the particles are allowed to move 
to a nearest neighbor site. 

In the pure diffusive case, no preferred direction for the jump is assigned.  
The reaction takes place only when  
 a certain number $k$ of the particles: $kA \rightarrow lA$ with $l<k$. 
The moves may take place using either  parallel or asynchronous dynamics. 
In the parallel dynamics scheme, the positions of all the particles are 
updated simultaneously while  
for the  asynchronous dynamics, the positions  of the particles 
are  updated  randomly one at a time.
With asynchronous dynamics, annihilating unbiased diffusive particles with $k=2$ and $l=0$ mimic the dynamics of voter models and 
the zero temperature Glauber-Ising model in one dimension while  coalescing random particles 
map onto the voter model in all dimensions. Such systems have been studied in one dimension 
\cite{books,spouge,amar,avraham1,alcaraz, schutz,krebs,racz,santos,sasaki,brunet,oliveira}
as well as in higher dimensions \cite{oliveira,cardy,kang,zumofen,peliti,droz,lee,cardy1,gyorgy}. 

The steady state of the process is 
rather simple. The number of particles left in the system at steady state 
will depend on the  dynamics. 
In asynchronous dynamics,
 $k=2$ always while  $l$ can be 1 or zero.  With $l =0$,  i.e., $A + A \to \emptyset$, the number of particles in the steady state will be zero or one for initial states having  even or 
odd number of particles respectively. 
For parallel dynamics,  
$k$ may be greater than 2. With $l=0$, the 
steady state  number  of particles is zero, one  or two \cite{two} and it does not  depend 
on the initial state.

The point of interest in all 
these analyses is how the system approaches the steady state. In particular, one intends to know 
how the number of particles decays with time. 

Another quantity of interest in this context is  the persistence probability $P(t)$, which 
accounts for the survival of a field up to time $t$. In many cases, persistence probability decays
as a power law, $P(t) \sim t^{-\theta}$, with the persistence exponent $\theta$ unrelated to 
other static and dynamic exponents \cite{majumder}.
For the  random walk problem or  reaction diffusion systems, $P(t)$ is defined 
as the probability that a site in the lattice remains unvisited
till time $t$ and is essentially related to first passage processes.  
For  the two dimensional unbiased annihilating reaction diffusion system 
  with asynchronous dynamics,  
 $\theta = 1/2$, 
obtained using  field theoretical
renormalisation  group
 method
\cite{cardy}. 
To the best of our knowledge $\theta$ for the purely diffusive case has been obtained only for asynchronous dynamics.


Several  systems with interacting entities  (e.g., bacteria and antibiotics,    predators and preys, individuals in a society etc.) can be studied using  reaction diffusion models with a bias.
 In the single species case, when the  particles
obey $A + A \to \emptyset$, if a   bias to move towards their nearest neighbours is considered, the system  has a direct mapping to a previously
proposed model for opinion dynamics \cite{BSR,pspr2015,biswas-sen}. One has to use asynchronous dynamics to preserve the mapping.
In this paper, we have considered this single species problem 
in two dimensions. 
Although the  mapping to a corresponding spin/opinion model is no longer obvious here, 
the  model may be  regarded as  a minimal model for herding behaviour  
where particles tend to form groups \cite{turchin}. This is mimicked by the attractive bias. 
On the other hand, the annihilation process may represent the fact that 
 the group `closes' and becomes dormant such that the particles forming the group are no 
longer under consideration.   The group size is just 2 for the asynchronous dynamics and 
can vary between 2 to 4 for parallel dynamics; the annihilation 
ensures that any  group of size greater than 1 is closed for both cases. 

As already mentioned,  often reaction diffusion  can be mapped to 
spin systems and an alternative way of investigating the spin systems 
is to study the corresponding reaction diffusion model with 
 asynchronous updating.
However, there is no harm in  considering the reaction diffusion system as an independent problem and one can use, in principle,  both asynchronous and parallel dynamics \cite{gameoflife}. In fact, it is of interest to study
 both the
nonequilibrium behaviour as well as the steady states for different dynamical 
schemes  as it has a potential application in other phenomena e.g., in naturally or artificially generated 
pattern formation. In the present paper, we consider parallel dynamics in general. 
 

%
To further generalise the problem, we also consider the bias to be either positive or negative.  
We are primarily interested in the time evolution of the density of particles $\rho(t)$, the nature of the steady states  and persistence probability $P(t)$.
The persistence behaviour is expected to be different from \cite{cardy} as parallel dynamics is used.   It is of interest whether the persistence exponents 
for the  asynchronous and parallel dynamics are related in the same way as was 
observed for 
  the Ising Glauber model \cite{derrida-exact,ray-mapping} and Potts model \cite{potts}.

%



\section{Dynamics and simulation methods}

We consider two types of dynamical schemes in the simulation. 
In the first case,  the particles have a bias to move towards their nearest neighbours 
with probability 
$\varepsilon$ and are diffusive otherwise (Case I). The unbiased case for $\varepsilon = 0$ has been  considered earlier \cite{kang,zumofen,peliti,droz,lee,cardy1,gyorgy}. In order to realise the case that the particles  strictly avoid the nearest neighbour, one can introduce a 
second type of dynamics (Case II). 
Here  the bias to move towards the nearest neighbour is $\varepsilon^\prime$ and with probability $(1-\varepsilon^\prime)$, they move to a  direction away from the nearest neighbour. 
One can establish a  relation between $\varepsilon$ and $\varepsilon^\prime$ in the following way:
 in Case I, the total probability of a particle to move one step towards its nearest particle is
$\varepsilon + \frac{1 - \varepsilon}{4}$, which for Case II, is equal to $\varepsilon^\prime$.
Therefore,
\begin{equation}
\varepsilon + \frac{1 - \varepsilon}{4} = \varepsilon^\prime.
\label{relation}
\end{equation}
Putting  $\varepsilon^\prime  = 1/4$ gives the diffusive (unbiased) limit in Case II, which, according to Eq. (\ref{relation}) correctly  corresponds to the diffusive case $\varepsilon = 0$ for Case I.
For the extreme bias limits, $\varepsilon = \varepsilon^\prime =1$, which is also obtained from Eq. (\ref{relation}). 

In the  simulation, one has to be careful as there may be more than one nearest particle. Also, in Case II,
when the move towards the nearest particle is not chosen, one has to make sure that the actual move does not
bring it closer to the nearest particle. This can happen when there are 
more than one particle which are at nearest distance  and/or  non-unique
 moves that brings it closer to the nearest particle. However, it may be expected that at long times when
the density of particles becomes small due to annihilation, the nearest particle will be unique and also the possibility
of non-unique moves bringing the particle closer to its nearest particle will become rare. As for larger system sizes the
dynamics continue for a long time, it is expected that in the simulation, the above equality will hold in the
thermodynamic limit.

Case II in principle includes Case I and allows both 
 positive and negative bias as desired. However, Case I being easier to handle computationally, 
we perform simulation for Case I for $0 \leq \varepsilon \leq 1$ and for Case II, mainly the region $\varepsilon^\prime \leq 0.25$, which cannot be realised in Case I. The results for  $ 0\leq \varepsilon \leq 1 $ and $\varepsilon^\prime \leq 0.25$ will constitute the entire regime of $\varepsilon^\prime$, i.e., Case II. 


The studies were performed on $L \times L$ square lattices with $L \le 256$ for Case I and $L \le 128$ for Case II (unless otherwise specified). Periodic boundary condition is
used in the simulations.
Initially the lattice is  randomly half-filled and the density of particles $\rho(t)$ is scaled by the initial occupation density  such that $\rho(0)=1$. 
In one Monte Carlo Step (MCS) all the particles are 
allowed to move simultaneously.
The annihilation takes place after the completion of one MCS.
The simulations are performed up to a maximum time of $10^7$ so that a
steady state is obtained within the maximum number of iterations (exceptional cases
have been discussed in detail in section IV). In the steady state
the number of remaining particles attain a constant value.
 During the time evolution, if the number of  remaining particles becomes  $0$ or $1$, the simulations are
terminated.

\section{Results for case I}

The case where the particles move towards their nearest neighbours with probability $\varepsilon$ and is diffusive otherwise is discussed in this section.
If two or more particles are found on the same site after the completion of one MCS, all of them are annihilated.  
The limit $\varepsilon =0$ corresponding to the purely diffusive case was considered earlier \cite{kang,zumofen,peliti,lee,cardy1,gyorgy}
and the variation of $\rho(t)$, the particle density,  was found from field-theoretic and numerical methods to be 
\begin{equation}
\rho(t) =\frac{ a + b \ln(t)}{t}.
\label{rhoeq}
\end{equation} 
In our simulation we recover the above variation with 
with $a = 0.6795 \pm 0.0014$ and $b = 0.6453 \pm 0.0002$
for $L = 512$, consistent with earlier estimates \cite{note}.
We have also checked that the values of $a$ and $b$ for asynchronous dynamics are considerably
different from the ones obtained for the parallel case, although the form remains same.

\begin{figure}
\includegraphics[width=9cm]{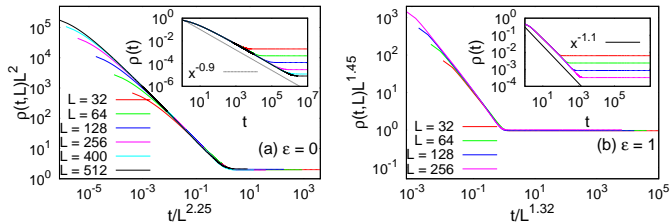}
\caption{Data collapses of $\rho(t,L)$ for (a) $\varepsilon = 0$ and (b) $\varepsilon = 1$. Insets show the unscaled data.} 
\label{rhoscale_t1}
\end{figure}

Eq. (\ref{rhoeq}) shows that one can approximately regard the variation of $\rho(t)$ as a power law decay.  The dominating term at large times being proportional to $\ln(t)/t$, the approximate value of the exponent  would be somewhat less than 1. 
Fig. \ref{rhoscale_t1}  shows that it is possible to collapse the data for different
system sizes with appreciable accuracy using suitable scaling factors.  
Assuming that $\rho(t)$ approximately decays as $t^{-\mu}$, 
the data collapse occurs with the following scaling form:
\begin{equation}
\rho(t,L) =  L^{-y} f(t/L^{z}).
\label{xyz1}
\end{equation}
To regain the form $\rho(t) \sim t^{-\mu}$ for large $L$, the scaling function $f(x)$ must behave as $x^{-\mu}$ such that 
\begin{equation}
y = z \mu.
\label{xyz0}
\end{equation}
The best collapse was  found with $y = 2$ and $z = 2.25$ for $\varepsilon =0$, giving $\mu \approx 0.9$ (Fig. \ref{rhoscale_t1}(a)).

\begin{figure}
\includegraphics[width=8cm]{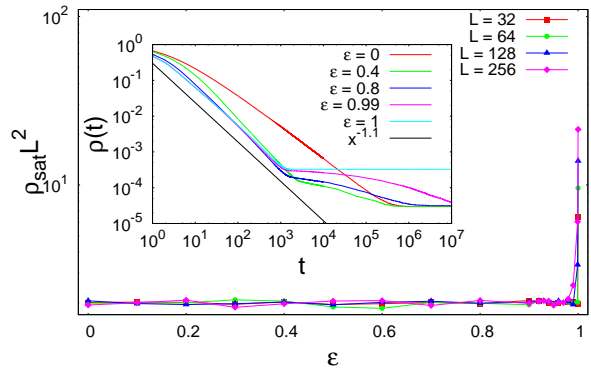}
\caption{$\rho_{sat} L^2 $ shown against $\varepsilon$. Inset shows the variation of $\rho(t)$ as a function of time $t$ for a system size of $256 \times 256$ for several values
of $\varepsilon$.} 
\label{satval1}
\end{figure}

As $\varepsilon$ is made non-zero, three different regions emerge in the behavior of $\rho(t)$. 
Initially there is a  fast decay followed by a slower decay and at later stages, a saturation region appears (inset of Fig. \ref{satval1}).
The intermediate region of slower decay is prominently observed for $0 < \varepsilon < 1$, but vanishes at $\varepsilon = 1$ and $0$.
The form of $\rho(t)$ for $\varepsilon =0$ given by Eq. (\ref{rhoeq}) is not seen to be followed for any nonzero value of $\varepsilon$.
Nevertheless, for $\varepsilon =1$, an analysis similar to that made for 
 $\varepsilon =0$  can be attempted to obtain an effective exponent.
We find a nice collapse  for the data for different system sizes  using the values 
 $y = 1.45$ and $z = 1.32$ (Fig. \ref{rhoscale_t1}(b)).
Thus the effective decay exponent $\mu$ for $\rho(t)$ is $\approx 1.1$ which is different from that of $\varepsilon =0$. In fact the initial faster decay region for all $\varepsilon >0$ appears to have an approximate power law behavior with the value of the exponent $\mu$ varying between 1.3 and 1.1 (shown in inset of Fig. \ref{satval1}). 
It is to be noted here that for $\varepsilon=0$ the estimate of $z$ can be 
made more accurate  by using the exact variation  of $\rho(t)$ (see section IV),
the present estimate is only to make a fair comparison with the $\varepsilon=1$ case.

We next study $\rho_{sat}$, the saturation value of the density of the particles   as a function of $\varepsilon$.
In section I it was mentioned that one expects three possibilities: zero, only one or two surviving  
particles at steady state even for the unbiased case.
This will make $\rho_{sat}$, the saturation value,  vanishingly small as the system size is increased. Indeed, for 
 $\varepsilon < 1$, $\rho_{sat}$  
is found to be  $ \ll 1$, vanishing as $1/L^2$. 
However for $\varepsilon = 1$, $\rho_{sat}$ is one order of magnitude higher than those for $\varepsilon < 1$ (inset of Fig. \ref{satval1}).
This indicates that   a much larger  number of particles remain in the system at the steady state for $\varepsilon = 1$.
Indeed, the total number of remaining particles
$\rho_{sat} L^2$ shows an abrupt increase
 at $\varepsilon = 1$ as shown in Fig. \ref{satval1}.
In fact  for  $ \varepsilon = 1$, $\rho_{sat}L^2$ increases with system size while  for $\varepsilon \neq 1$, $\rho_{sat} L^2$ is fairly  a constant.

\begin{figure}
\includegraphics[width=9cm]{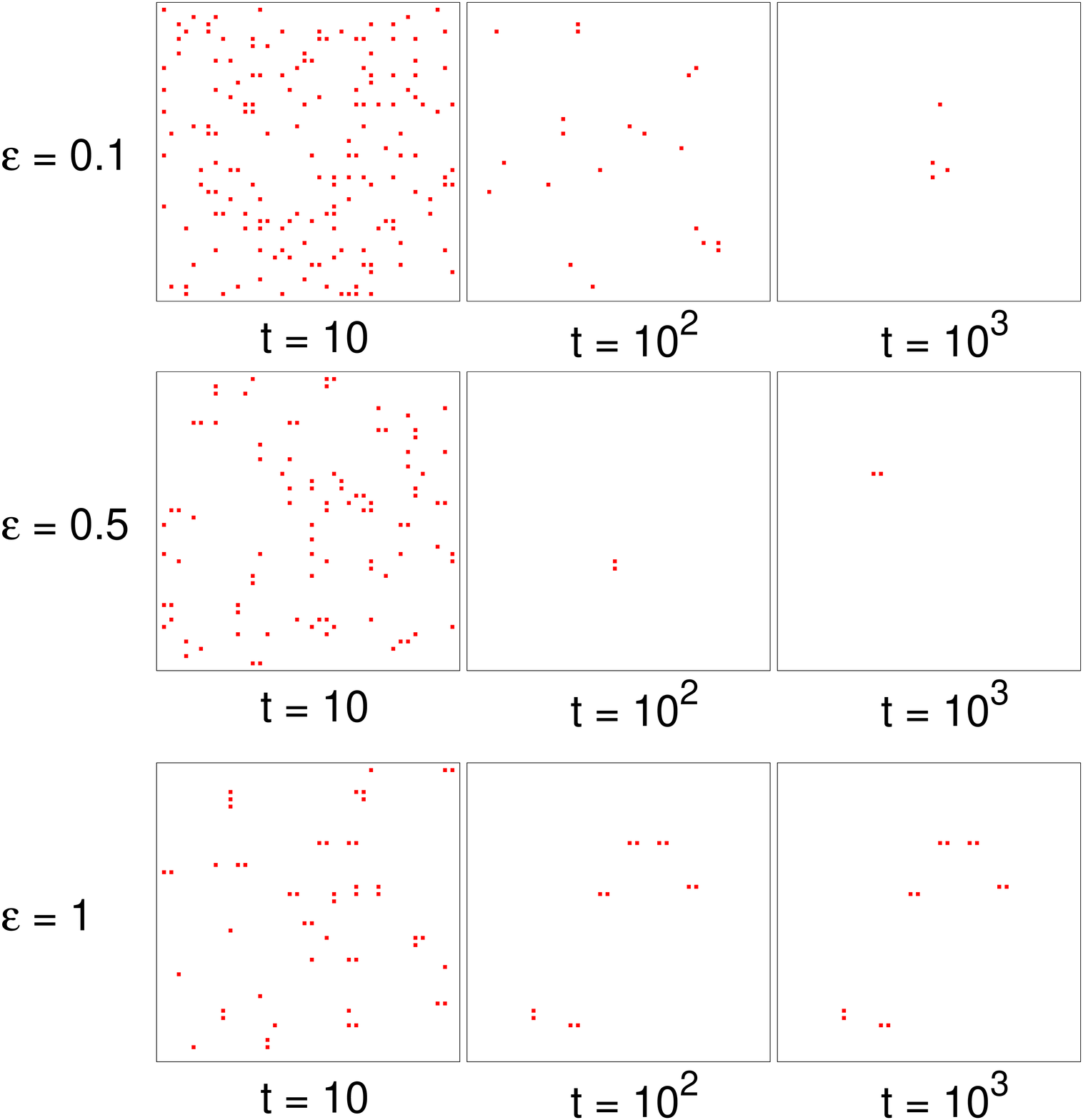}
\caption{Snapshots of a $40 \times 40$ system for $\varepsilon = 0.1, 0.5$ and $1$ at $t = 10, 10^2$ and $10^3$.
Cases for $\varepsilon = 0.5$ and $1$ shows the formation of dimers.
For $\varepsilon = 1$ the dimers are seen to persist.} 
\label{snap}
\end{figure}

\begin{figure}
\includegraphics[width=8cm]{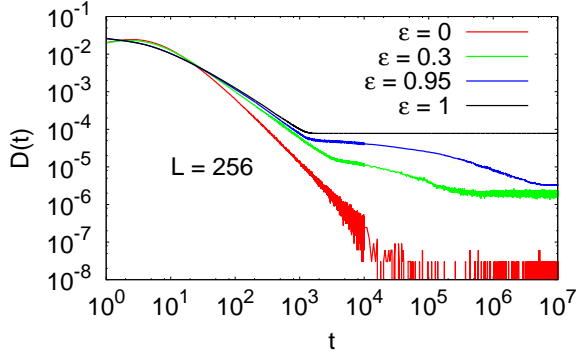}
\caption{Variation of $D(t)$ as a function of time $t$ for several $\varepsilon$ values for system size of $256 \times 256$.}
\label{dtvst}
\end{figure}

At  $\varepsilon =1$,  it can be easily understood that at long times when few particles are left, as two particles always approach each other, a situation will emerge when they may appear on neighbouring sites with no other neighboring sites occupied. Fig. \ref{snap} shows the snapshots
at different times for different values of $\varepsilon$ and at large times for $\varepsilon = 1$, one notes that such pairs will survive indefinitely. This is because 
the subsequent dynamics will simply be a swapping of the positions of the two neighbouring particles. These two particles  form what we call a dimer. 
A  dimer is defined to be a pair of particles occupying neighbouring sites and both the particles do not have any other neighbour. 
As these particles will never get annihilated at $\varepsilon = 1$, it is expected that  there will be a larger number of surviving particles. To probe this 
in more detail, we calculate 
 $D(t)$, 
the density of dimers in the system
for all $\varepsilon $ values. 
$D(t)$ is the number of dimers divided by $L^2$.
The data shown in Fig. \ref{dtvst} clearly show a larger value of  $D(t)$ for $\varepsilon =1$ in the steady state compared to other $\varepsilon$ values as expected. However, surprisingly, a non-zero saturation value is found for other values of $\varepsilon > 0$ also.
At $\varepsilon = 0$, the value of $D(t)$ at large times becomes negligibly small ($\mathcal{O}(10^{-8})$).
This is expected as even though two particles can remain in the steady  state 
the probability of
forming dimers for $\varepsilon = 0$ is negligible, as the particles have equal probability
of moving to other neighboring sites.

\begin{figure}
\includegraphics[width=9cm]{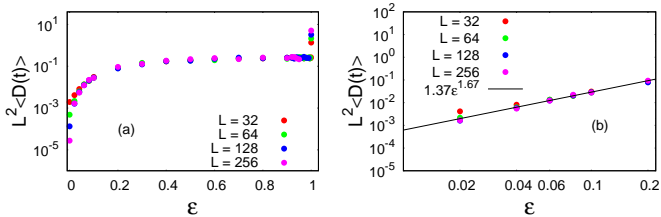}
\caption{(a) Variation of $\langle D(t) \rangle$ in the steady state multiplied by $L^2$ with $\varepsilon$ for several system sizes. (b) Power law behavior of this quantity close to $\varepsilon = 0$.}
\label{dim22}
\end{figure}

Studying the saturation values of the actual number of dimers, $\langle D(t) \rangle L^2$ ($\langle ... \rangle$ indicates a thermal average in the steady state) against $\varepsilon$ (Fig. \ref{dim22}(a)), we note that it has a nonzero finite value increasing with $\varepsilon$ (independent of $L$ for $\varepsilon < 1$) and a jump at $\varepsilon = 1$ (increasing with $L$).
Close to $\varepsilon = 0$, $L^2\langle D(t) \rangle$  shows a power law behavior with $\varepsilon$ in the form $\sim \varepsilon^{\eta}$ with $\eta = 1.67 \pm 0.04$ (Fig. \ref{dim22}(b)).

For nonzero $\varepsilon$ therefore, the steady state can have several 
possibilities:  (i) zero,   one, or    two isolated particles (ii) dimers (iii) coexistence 
of dimers and isolated particles.
For $\varepsilon = 1$, the last case will not arise. Let us, for the sake of argument, assume that it is possible to have a state 
comprising of a dimer and an isolated particle. The dimer will remain static in its position, while the
isolated particle will travel to form a ``trimer", which is unstable and will decay immediately into  an isolated particle.
When $N_p(t)$, the  number of particles present at time $t$ is even, the maximum number of dimers that can be formed is $N_p(t)/2$.
We hence define,
\begin{equation}
d_r(t) = \frac{2D(t)L^2}{N_p(t)}
\label{ratio}
\end{equation}
as the probability that a dimer is formed  out of the remaining particles.
We note that $d_r(t) \rightarrow 1$ in a finite time for $\varepsilon = 1$.
Therefore, if an even number of particles survive for $\varepsilon = 1$, all of them will form dimers.
Of course,  states with either zero or a single surviving particle are also possible for $\varepsilon =1$.
These features have all been checked in the simulation.

\begin{figure}
\includegraphics[width=9.1cm]{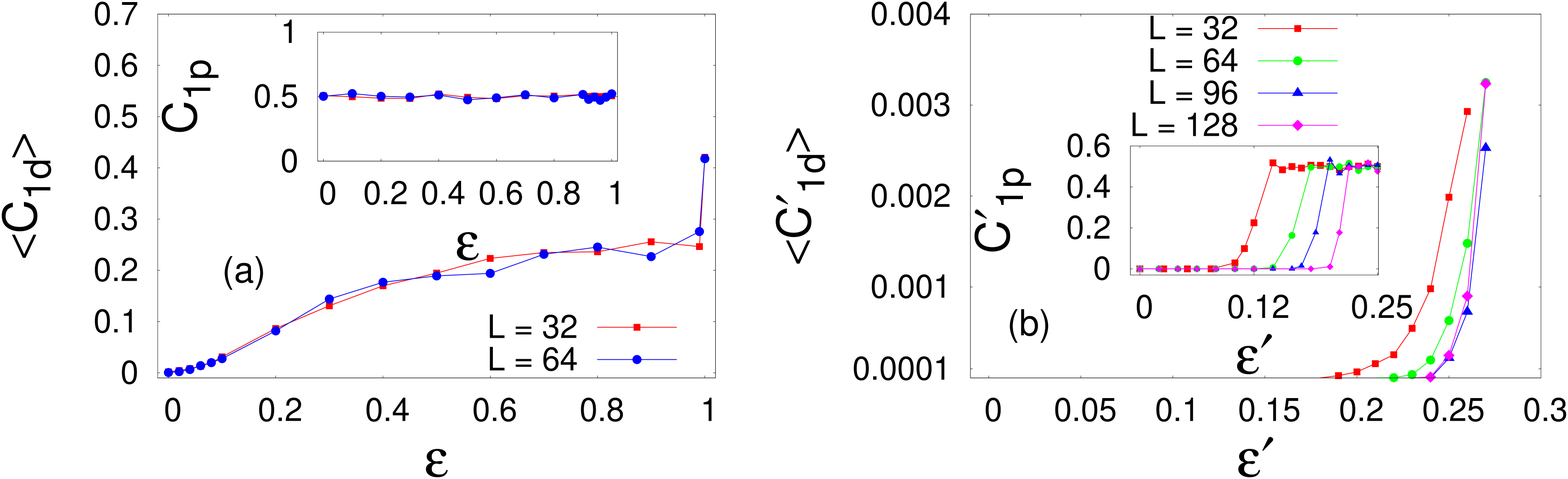}
\caption{Variation of (a) $\langle C_{1d} \rangle$ and $C_{1p}$ with $\varepsilon$ and that of (b) $\langle C'_{1d} \rangle$ and $C'_{1p}$ with $\varepsilon^\prime$.}
\label{1w1d}
\end{figure}

To get an idea which kind of steady states are more prevalent, we calculate $C_{1p}$, 
the fraction of cases 
for which the system is left with a single 
particle. $C_{1p}$ is found to be nearly 0.5 for all $\varepsilon$ (inset of Fig. \ref{1w1d}(a)).  We also calculate
$\langle C_{1d} \rangle$, 
 the probability 
of getting at least one dimer which 
turns out to be an increasing function of $\varepsilon$, showing a jump at $\varepsilon =1$ (Fig. \ref{1w1d}(a)).  Although the probability is less than 0.5, still 
it is quite appreciable ($\mathcal{O}(10^{-1}$)).

\begin{figure}
\includegraphics[width=8cm]{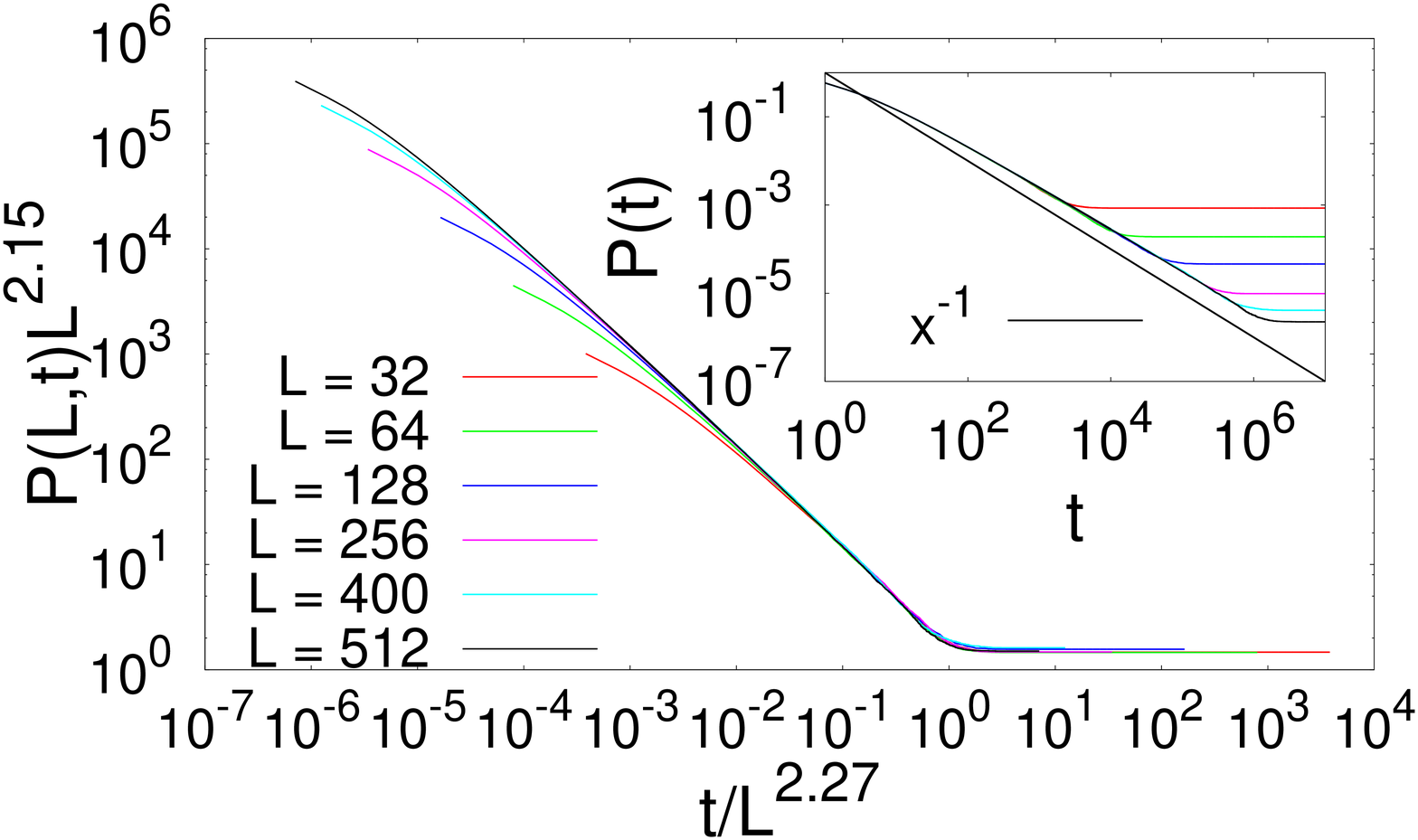}
\caption{Data collapse of $P(t)$ for $\varepsilon = 0$ using Eq. (\ref{per_scale_t1}). Inset shows the unscaled data.} 
\label{per_t1_e0}
\end{figure}

To calculate the persistent probability $P(t)$, the initially occupied 
sites are regarded as unvisited.
For $\varepsilon = 0$ the persistence probability $P(t)$ was calculated for a maximum size $L = 512$ and it shows a power law decay: 
$P(t) \sim t^{-\theta}$ with
$\theta \simeq 1$ (inset of Fig. \ref{per_t1_e0}). This value of the persistence exponent obtained for parallel dynamics,
is also, therefore, approximately twice the value for asynchronous dynamics, as was found in  the  one dimensional spin models \cite{ray-mapping,potts}. 
 Finite size scaling analysis was  performed using the scaling ansatz \cite{manoj}
\begin{equation}
P(t,L)L^{\alpha}=g(t/L^{z_p}),
\label{per_scale_t1}
\end{equation}
and a data collapse was obtained
with $\alpha = 2.15$ and $z_p = 2.27$ 
such that
$\theta = \alpha/z_p \approx 0.95$.
The results are shown in Fig. \ref{per_t1_e0}. 
As $\varepsilon$ is increased, $P(t)$ deviates from the power law
behavior and saturates to higher values. Since annihilations are enhanced for nonzero $\varepsilon$, a higher saturation value is obtained as $\varepsilon$ is increased.  

\section{Results for case II}

In the simulation for Case II dynamics, a particle is allowed to move one step towards its nearest particle  with probability $\varepsilon ^\prime$
and with probability $1 - \varepsilon ^\prime$ it goes to one of the three remaining nearest neighboring positions. The annihilations
take place in the usual manner.
The region $0.25 < \varepsilon^\prime \le 1$ of Case II corresponds to $0 < \varepsilon \le 1$ of Case I, where the
particles have the tendency to approach their nearest particle on the lattice, i.e., the bias is positive.

We first check the validity of Eq. (\ref{relation}) using the Case II dynamics; specifically it is verified where the 
 diffusive point lies. Here it should be mentioned that when the particle does not choose to move to a site that brings it closer to 
its nearest particle, it moves to one of the  remaining three neighbouring sites. This implies that the fact that one of these movements may also take it closer to its nearest particle is ignored. Hence one can expect a validity 
of Eq. (\ref{relation}) only in the thermodynamic limit, as argued in section II. 

As the annihilations become less probable for negative bias, the system 
takes a very long time to attain the steady state and in the simulations, for larger system sizes, the  system may not attain the steady state within the maximum number of Monte Carlo steps. 
Nevertheless, some essential information could  be extrapolated from 
the results obtained in the nonequilibrium region. 
	
At $\varepsilon ^\prime =0$, the particles are completely repulsive and even at very long times, one may expect a 
nonzero density of particles for small $\varepsilon ^\prime$. 
At the diffusive limit on the other hand, the number of remaining particles being either zero, 1 or 2, $\rho(t \to \infty) \to 0$. It is also highly unlikely 
that dimers will be formed with the negative bias and one expects the number of configurations with dimers to be equal to zero till the diffusive point. 
As in Case I, here too we estimate   $C_{1p}^\prime$ and $\langle C_{1d}^\prime \rangle$, the probabilities of getting a single surviving particle
and  at least one dimer respectively. The results are shown in Fig. \ref{1w1d}(b). $\langle C_{1d}^\prime \rangle$ clearly shows an abrupt
rise close to $\varepsilon' = 0.25$ as the system size increases, which supports the fact that the diffusive point lies at
$\varepsilon' = 0.25$. $C_{1p}^\prime$ also shows an abrupt rise, however, here the finite size effects
are stronger and the abrupt rise takes place at a value of $\varepsilon'$ which 
slowly  approaches 0.25. Hence the diffusive point, 
 denoted  by ${\varepsilon_c}' = 0.25$  in the thermodynamic limit, depends on the system size. 
 A detailed study of $\rho(t)$ given in the following indicates that the remaining fraction of particles at very large times varies as  
$({\varepsilon_c}' - \varepsilon')^{\beta}$, where  
${\varepsilon_c}'=0.25$
and $\beta\approx 2$. 
The sharp decrease of $\rho$ for $\varepsilon'$ close to ${\varepsilon_c}'$ 
implies  there could be     a  finite number of configurations having only one remaining particle 
even for $\varepsilon' < 0.25$ and this explains why $C_{1p}^\prime$ shows the
abrupt increase at a lower value of  $\varepsilon'$ compared to 0.25,  even for the maximum size considered.

\begin{figure}
\includegraphics[width=9cm]{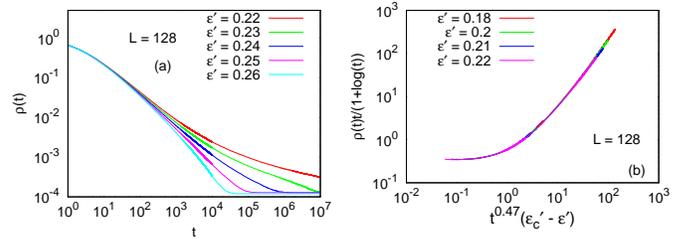}
\caption{(a) Variation $\rho(t)$ as a function of time $t$ for $L = 128$ for
several values of $\varepsilon ^\prime$. (b) Data collapse of $\rho(t)$ for $L=128$ for several values of $\varepsilon'$ using the scaling form Eq. (\ref{rhocol1}).}
\label{fract2}
\end{figure}

\begin{figure}
\includegraphics[width=8cm]{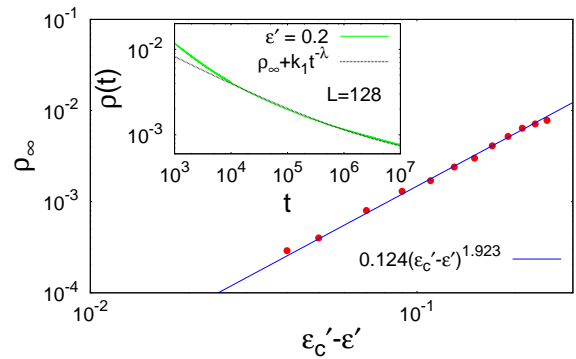}
\caption{Variation of $\rho_{\infty}$ as a function of ${\varepsilon_c}'-\varepsilon'$ for $L = 128$. The data was fitted according to the function 0.124$({\varepsilon_c}'-\varepsilon')^{\beta}$, with $\beta = 1.923 \pm 0.05$. Inset shows the fitting of a $\rho(t)$ curve for $L = 128$, $\varepsilon'=0.2$ according to the function $\rho_{\infty}+k_1t^{-\lambda}$, with $\rho_{\infty} = 0.004\pm10^{-6}$, $k_1 = 0.0819\pm10^{-4}$ and $\lambda = 0.34\pm10^{-4}$.}
\label{rhoatinf}
\end{figure}

\begin{figure}
\includegraphics[width=9cm]{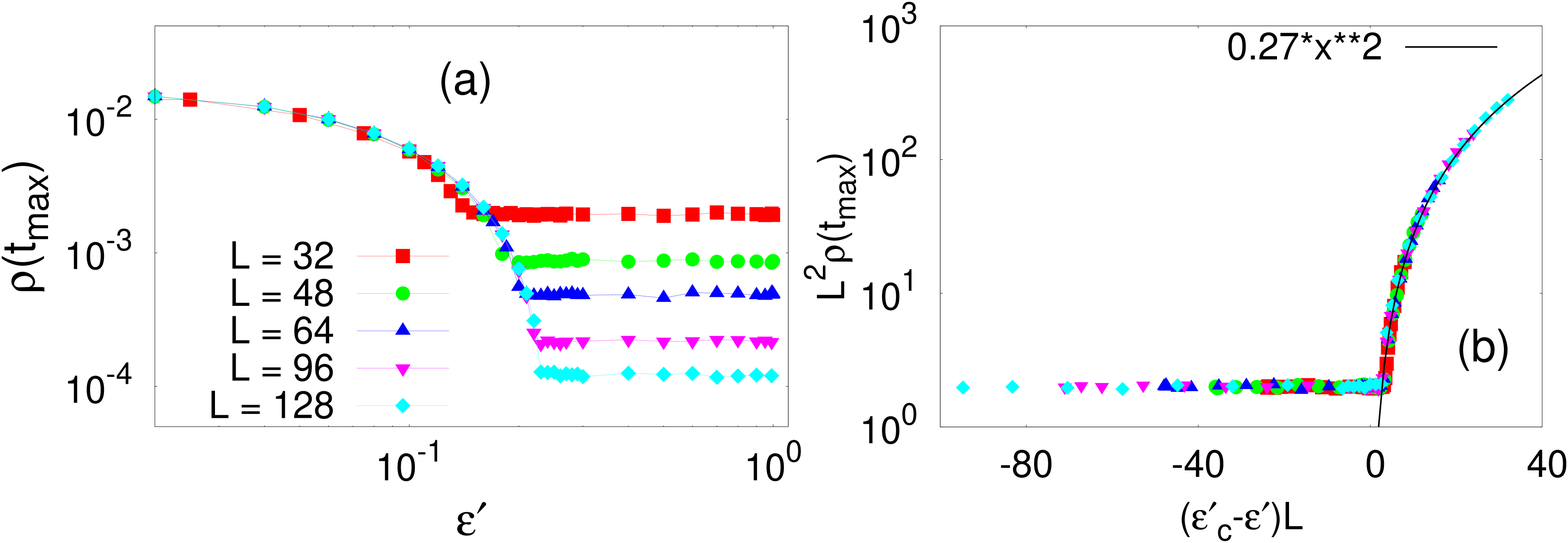}
\caption{(a) Variation of $\rho$ at $t_{max} = 10^7$ as a function of $\varepsilon'$. $\rho(t_{max})$ reaches a plateau at ${\varepsilon_c}'(L)$. (b) Data collapse of $\rho(t_{max})$ for several system sizes using the scaling relation given by Eq. (\ref{rhocol2}). The scaled form varies as $({\varepsilon_c}'-\varepsilon')^2$ for $\varepsilon'<{\varepsilon_c}'$ (shown by the solid line).}
\label{rhocol}
\end{figure}

\begin{figure}
\includegraphics[width=9cm]{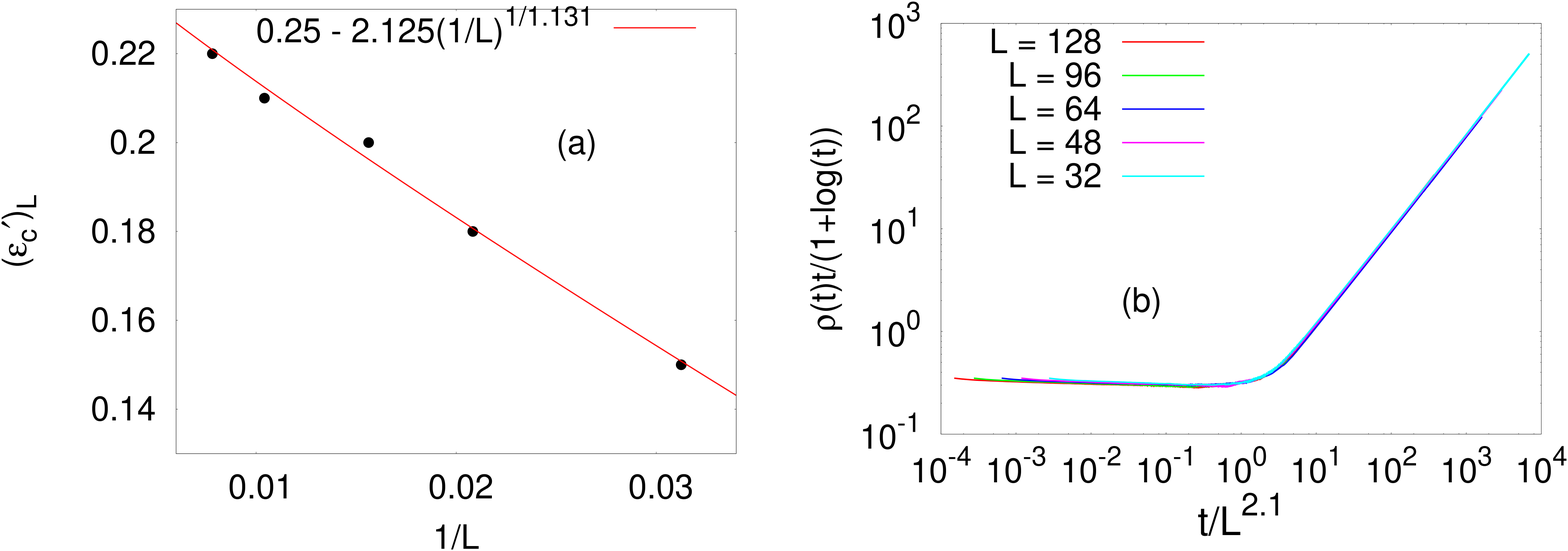}
\caption{(a) Variation of $({\varepsilon_c}')_{eff}$ with $1/L$. The data is fitted according to the curve $0.25 - 2.125x^{1/1.131}$. (b) Data collapse of $\rho(t)$ for several system sizes at $\varepsilon' = 0.25$ using the scaling form Eq. (\ref{rhocol1}).}
\label{epscvsL}
\end{figure}

Next, we analyse the behaviour of  $\rho(t)$ in detail. 
For the largest size simulated, $L = 128$, the data are shown in Fig. \ref{fract2}(a) and fitted to the form
\begin{equation}
\rho(t) = \rho_{\infty}+k_1t^{-\lambda}
\label{eq-rho}
\end{equation}
(inset of Fig. \ref{rhoatinf}). We find a good consistency in the fittings for $t>10^4$.
It is found that 
\begin{equation}
\rho_{\infty}\sim({\varepsilon_c}' - \varepsilon')^{\beta},
\end{equation}
 with $\beta = 1.92\pm0.05$ (Fig. \ref{rhoatinf}), while $\lambda$ increases with $\varepsilon'$, since convergence is reached faster for larger value of $\varepsilon'$.



The above result is also consistent with a finite size analysis when $\rho(t_{max})$ 
is studied  for different values of $L$ and one can argue that $\beta \approx 2$. Here, $t_{max}$ is the maximum number of Monte Carlo steps up to which the simulations were performed.
The unscaled data is shown in Fig. \ref{rhocol}(a).  We find the following scaling behaviour of $\rho(t_{max})$:
\begin{equation}
\rho(t_{max})L^s = f_1\Big(({\varepsilon_c}'-\varepsilon')L^{\alpha'}\Big),
\label{rhocol2}
\end{equation}
with $s = 2 $ and $\alpha'=1$ (Fig. \ref{rhocol}(b)). 
The scaling function  behaves as 
\begin{equation}
f_1(x) = 
\begin{cases}
{\rm const}  & \text{for $\varepsilon'>{\varepsilon_c}'$}\\
x^2 & \text{for $\varepsilon'<{\varepsilon_c}'$}.
\end{cases}
\end{equation}
This implies that $\rho(t_{max})$ varies as $L^{-2}$ for $\varepsilon'>{\varepsilon_c}'$ and as $({\varepsilon_c}'-\varepsilon')^2$ for $\varepsilon'<{\varepsilon_c}'$.
The scaling variable being $({\varepsilon_c}'-\varepsilon')L$, we conjecture that there is a diverging length scale 
$L\sim({\varepsilon_c}'-\varepsilon')^{-1}$. 
In support of this, we study the variation of  the effective critical value  ${\varepsilon_c}'(L)$,  determined by the value of $\varepsilon' $ at which $\rho$ attains a constant value 
as a function of $\varepsilon'$ 
(shown in Fig. \ref{rhocol}(a)) for the system size $L$.
$\varepsilon_c'(L)$ is found to obey the familiar scaling form for a continuous phase transition 
\begin{equation} 
\varepsilon_c'(L) = \varepsilon_c'-{\mathcal O}(\frac{1}{L^{1/\nu}}),
 \end{equation}
shown in   Fig. \ref{epscvsL}(a).
 Here, $\nu = 1.13
\pm0.06$, which is close to 1, consistent with the finite size scaling analysis of $\rho(t_{max},L)$.
We conclude  that $\rho$ behaves like an order parameter that vanishes at the critical point $\varepsilon_c'$ with the critical exponent values $\beta \approx 2$ and $ \nu \approx 1$.

The presence of a diverging time scale can also be detected  when we note that 
a data collapse for $\rho(t)$  for $L = 128$ is  obtained in the   following form
\begin{equation}
\rho(t)\Big(\frac{1+\log(t)}{t}\Big)^{-1} = f_2\big(t^{a_1}({\varepsilon_c}'-\varepsilon')\big)
\label{rhocol1}
\end{equation}
with $a_1 = 0.47\pm0.04$ (shown in Fig. \ref{fract2}(b)). This shows that the diverging time scales as $({\varepsilon_c}'-\varepsilon') ^{1/a_1}$. Hence, the dynamic exponent is $1/a_1 \approx 2.1$.
Also, we have been able to conduct a finite size scaling analysis of $\rho(t)$ 
exactly at criticality for different system sizes using the form
\begin{equation}
\rho(t)\Big(\frac{1+\log(t)}{t}\Big)^{-1} = f_3(t/L^{z}). 
\label{rhocol3}
\end{equation}
Once again we find a good collapse with $z \approx 2.1$. This result is   
shown in   Fig. \ref{epscvsL}(b).
It is to be noted that in Eqs. (\ref{rhocol1}) and (\ref{rhocol3}) we have used the exact form of $\rho(t)$ for $L\to\infty$.



\begin{figure}
\includegraphics[width=9cm]{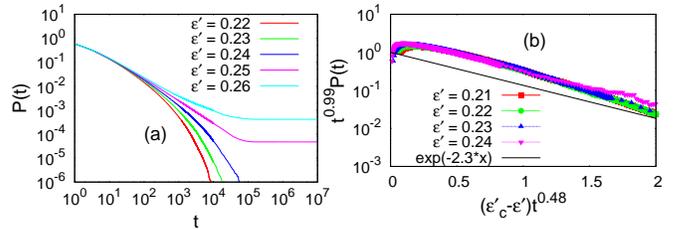}
\caption{(a) Variation of persistence probability $P(t)$ as a function of time $t$ for a $128\times128$ system for several values of $\varepsilon'$. (b) Data collapse of $P(t)$ using the scaling form given by Eq. (\ref{perst}). The collapsed form shows a variation of the form $\exp({-t/\tau})$.}
\label{pert2}
\end{figure}

Finally, the  persistence probability $P(t)$ is studied; for small $\varepsilon ^\prime$ it  shows a fast decay and goes to zero.
Above a certain value of $\varepsilon^\prime$, $P(t)$ saturates to a non-zero value. From Fig. \ref{pert2}(a) we can see that $P(t)$ changes in nature  at $\varepsilon ^\prime \approx 0.25$. This is consistent with the fact that in the 
small  $\varepsilon ^\prime$ region the particles have a tendency to move away 
from its nearest walker, thereby reducing the probability of annihilation  and allowing the
particles  to roam around more. For $\varepsilon^\prime > 0.25$, the particles annihilate at a faster rate, such that  a large number of sites can 
remain unvisited, as already discussed in section III.

For the system size $L =128$, the data for $P(t)$ are collapsed according to the following form,
\begin{equation}
P(t)=t^{-0.99}f'\Big(t^{0.48}({\varepsilon_c}'-\varepsilon')\Big).
\label{perst}
\end{equation}
This has been shown in Fig. \ref{pert2}(b).
$f'$ shows an exponential decay for large values of 
the argument  such  that 
for large $t$, $P(t)$ varies as $t^{-0.99}\exp\big(t^{0.48}({\varepsilon_c}'-\varepsilon')\big)$. This form is consistent with the power 
law behaviour at the critical point as obtained in the previous section.

\begin{figure}
\includegraphics[width=8cm]{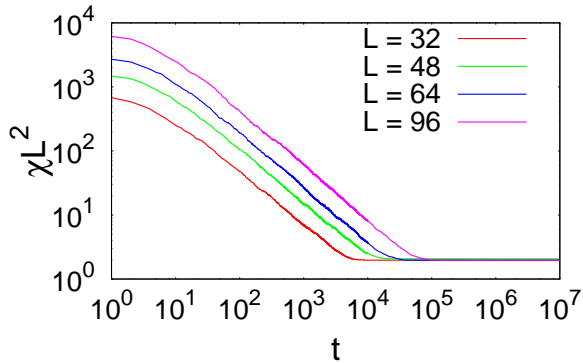}
\caption{Scaling collapse of the steady state values of susceptibility $\chi$ for several system sizes at the critical point ${\varepsilon_c}' = 0.25$. }
\label{susc}
\end{figure}

\section {Discussions and concluding remarks}

In this paper we have studied the  time evolution of a single species reaction diffusion system where the 
particles perform nearest neighbour hops on a lattice with a bias to move towards the nearest particle. The bias can be 
made both positive and negative (Case II) parametrically. The particles are annihilated if they are 
found on the same site after the completion of one Monte Carlo step. 
A  diffusive motion occurs for individual particles when the  parameter $\varepsilon^\prime = 0.25$, above (below) this
value, the particles have attractive (repulsive)  motion.

The dynamical evolution of the density of particles $\rho(t)$ is studied and as expected, for  $\varepsilon^\prime = 0.25$, $\rho(t)$ vanishes in the thermodynamic 
limit as the number of surviving particles can be zero, one or two. 
The region where $\rho(t \to \infty)$ remains finite is $\varepsilon' < \varepsilon'_c =0.25$;
the extrapolated data  for the maximum size simulated as well as a finite size scaling analysis indicate $\rho(t\to\infty)$ varies as $({\varepsilon_c}'-\varepsilon')^2$.


 Although $\rho(t) $  is still vanishingly small for  $\varepsilon^\prime > 0.25$, the decay of $\rho(t)$ takes place in a manner distinctly different from that of $\varepsilon^\prime =  0.25$. Moreover, 
a discontinuity occurs at  $\varepsilon^\prime = 1$, the fully positively biased point where the particles invariably move towards the nearest
particles. On probing deeper, we find a dimerisation taking place for any  $\varepsilon^\prime > 0.25$,  the number of  dimers as well as the probability  shows a jump at  $\varepsilon^\prime = 1$. The variation of the number of dimers (although its density is  vanishingly small)   
shows a  continuous increase from zero at the diffusive point; 
$\langle D(t)\rangle L^2 \propto (\varepsilon^\prime - 0.25)^{\eta}$, where $\eta$ is very close to 5/3.


The diffusive behaviour therefore occurs at a single point of the parameter space. This is reminiscent of the generalised voter model in two dimensions where
the voter model occurs at a single point of the two dimensional parameter space \cite{gen-voter}. 
The diffusive point  separates  two regions marked by completely different features: to its left one has a  nonzero value of $\rho(t \to \infty)$   and to the right,  
a nonzero number of dimers, both vanishing in  a  power law manner with different  exponents. 
The persistence probability also shows different behaviour 
for $\varepsilon^\prime < 0.25$ and $\varepsilon^\prime > 0.25$.
For $\varepsilon^\prime < 0.25$, it goes to zero in a stretched exponential manner while in the other region it attains a nonzero value at large times for $\varepsilon^\prime > 0.25$. Exactly at the diffusive point it shows a power law decay. 


We have concluded that a continuous   phase transition 
takes place  at 
$\varepsilon'=0.25$.   $\rho(t \to \infty)$ behaves  as an order 
parameter that 
  vanishes continuously  as $(\varepsilon_c'  - \varepsilon^\prime)^\beta$. The  phase transition is similar to that studied recently in \cite{daga}.
The presence of a diverging length scale as well as a diverging   timescale have  been 
detected as  ${\varepsilon_c}' \to 0.25$ with exponents $\nu \approx 1$ and $z$ lying between 2.1 and 2.2, obtained using several different finite size scaling analyses.

One interesting point to be noted is that as $\nu \approx 1$, $\alpha$, 
the critical exponent for the specific heat must be zero according to the scaling relation $\nu d = 2 - \alpha$ as $d=2$. In order to satisfy the scaling relation $\alpha + 2\beta + \gamma$ =2, where $\gamma$ is the 
critical exponent for susceptibility, $\gamma$ has to be negative. Indeed, we find that the steady state value of the susceptibility defined as $\chi = \big(\langle\rho^2\rangle-\langle\rho\rangle^2\big)L^2$ \cite{gyorgy}, varies as $L^{-2}$ such that $\gamma = -2\nu \approx 2$ (shown in Fig. \ref{susc}). The susceptibility 
not showing a divergence at the critical point is a counter-intuitive result, 
however, such a behaviour is not unprecedented and was  also 
observed for the dynamical models studied in  
 \cite{lalla,psen}. The anomalous behaviour of the susceptibility    could be due to the fact that the system
here involves long range interactions which can affect the values of the 
exponents non-trivially.


In general, asynchronous and parallel dynamics are expected to affect the dynamics differently; however, at least for the diffusive limit
$\varepsilon^\prime = 0.25$, the form of $\rho(t)$ is identical for both types of dynamics. Interestingly,   the persistence exponent for the parallel dynamics is found to be twice of that for the asynchronous, a result similar to the case for one dimensional spin models but by no means obvious.
The formation of the dimer patterns is an artifact of the parallel updating scheme.


Acknowledgement: This work was inspired by a comment made by Subir K. Das. 
Useful discussions with Purusattam Ray and Bikas Chakrabarti are acknowledged. P. Mullick thanks DST-INSPIRE (Sanction No. 2015/IF0673) for their financial support.
P. Sen acknowledges SERB (Scheme no EMR/2016/005429, Government of India) for  financial grant.


\begin{thebibliography}{99}
\bibitem{two}
Two  particles will survive in the parallel dynamics  for a special case only.
One can  regard the two dimensional lattice as a checkerboard with the particles occupying either black or white squares.  If the last two 
survivors  are on different coloured squares, in a parallel dynamics they will never meet and annihilate. 
However for asynchronous dynamics, they will be  annihilated ultimately, 
even if it takes a long time.
\bibitem{note}
In \cite{gyorgy}, Eq. (\ref{rhoeq}) was written in terms of
$\log_{10}t$ instead of natural $\log$. Making the necessary conversion,
the value of $b$ is found to be very close.
\end{thebibliography}
\end{document}